\documentclass{ws-rv975x65}
\usepackage{ws-rv-van}             % numbered citation/references (default)
\usepackage{ws-index}             % to produce multiple indices
\usepackage{epstopdf}
\makeindex
\newcommand{\be}{\begin{eqnarray}}
\newcommand{\ee}{\end{eqnarray}}

\def\pr{Phys. Rev.}

\def\la{\langle}\def\ra{\rangle}
\def\del{\partial}
\def\calL{\cal L}

\let\emph=\relax

\let\mathbf=\boldsymbol
\def\beginABC{\begin{subequations}}
\def\endABC{\end{subequations}}
\def\Tr{\rm Tr}
\def\bi{\bibitem}

\def\be{\begin{eqnarray}}
\def\ee{\end{eqnarray}}

\def\roughly#1{\mathrel{\raise.3ex\hbox{$#1$\kern-.75em%
\lower1ex\hbox{$\sim$}}}}
\def\lsim{\roughly<}
\def\gsim{\roughly>}
\def\la{\langle}
\def\ra{\rangle}
\def\Tr{\rm Tr}

\def\su2hls{HLS$_{SU(2)}$}
\def\u2hls{HLS$_{U(2)}$}
\def\u1hls{HLS$_{U(1)}$}
\def\O{\cal O}

\begin{document}

\centerline{\large\bf Fractionized Skyrmions in Dense Compact-Star Matter\footnote{To appear in {\it The Multifaceted Skyrmion} (2nd Edition) (World Scientific, Singapore)
ed. by M. Rho and I. Zahed}}
\vskip 0.5cm

\author{Masayasu Harada$^a$,  Hyun Kyu Lee$^b$, Yong-Liang Ma$^c$ and Mannque Rho$^d$}
%\vskip 0.5cm
\address{$^a$Department of Physics,  Nagoya University, Nagoya, 464-8602, Japan \\ harada@hken.phys.nagoya-u.ac.jp}
\address{$^b$Department of Physics, Hanyang University, Seoul 133-791, Korea\\ hyunkyu@hanyang.ac.kr}
\address{$^c$Center of Theoretical Physics and College of Physics, Jilin University, Changchun, 130012, China\\ yongliangma@ilu.edu.cn}
\address{$^d$Institut de Physique Th\'eorique, CEA Saclay, 91191 Gif-sur-Yvette, France\\ mannque.rho@cea.fr}

\begin{abstract}

The hadronic matter described as a skyrmion matter embedded in an FCC crystal is found to turn into a half-skyrmion matter with vanishing (in the chiral limit) quark condensate and {\it non-vanishing} pion decay constant {$f_\pi$} at a density $n_{1/2}$ lower than or near the critical density $n_c$ at which hadronic matter changes over to a chiral symmetry restored phase with possibly deconfined quarks. When hidden local gauge fields and a dilaton scalar of spontaneously broken scale symmetry {with decay constant $f_\chi$} are incorporated, { this half-skyrmion phase} is characterized by $f_\chi\approx f_\pi\neq 0$ with the hidden gauge coupling $g\neq 0$ but $\ll 1$. While chiral symmetry is restored {\it globally} in this region in the sense that space-averaged, $\la\bar{q}q\ra$ vanishes, quarks are still confined in massive hadrons and massless pions. This phase is shown to play a crucial role in the model for a smooth transition from a soft EoS at low density to a stiffer EoS at high density, the changeover taking place at $n_{1/2}$. It resembles the ``quarkyonic phase" predicted in large $N_c$ QCD and represents the ``hadronic freedom" regime which figures as a doorway to chiral restoration. The fractionization of skyrmion matter into half-skyrmion matter has a tantalizing analogy to what appears to happen in condensed matter in (2+1) dimensions where  half-skyrmions or ``merons" enter as relevant degrees of freedom at the interface.

\end{abstract}
%\smalltoc
\newpage

\tableofcontents
\setcounter{footnote}{0}
%\body
\section{Introduction}

Fractionized solitons play a singularly intriguing and sometimes crucial role in various areas of physics. Particularly striking is the impact of fractional solitons, such as half-skyrmions, merons, dyons etc. in highly correlated condensed matter systems and in holographic dual picture of baryons as amply illustrated in this volume.  Their role in hadronic matter has been, on the contrary, little if at all explored and much less is understood but the initial effort made in this direction shows a promising avenue. Two of us (HKL and MR) contributed a chapter on the past development  in the first volume of ``The Multifaceted Skyrmion" and since then a great deal of progress has been made, with deeper understanding of the basic mechanism involved. In this note, we update that contribution, improve on the conceptual aspect on the role of the dilaton and hidden local fields living with the pionic fields that provide the topology of the soliton.

In this article, we focus on the role and impact of the topology change from skyrmions to half-skyrmions in dense baryonic matter, considered to be relevant to the interior of compact stars. Unlike in condensed matter systems, effects of fractionized solitons cannot be directly exposed from measurable observables of correlated baryonic systems. Hence they can only be inferred indirectly from measured quantities. In the case we are concerned with, the system is driven to high density either by strong interactions or by gravity and the relevant quantity is the equation of state (EoS for short)  of neutron-rich dense matter appropriate for compact stars. This is the area that is currently the most challenging in nuclear physics.

It should be stressed at the outset that the approach anchored on topology in nuclear physics, although initiated by Skyrme in early 1960's for nuclear physics, is an approach barely explored and largely unfamiliar to most of the workers in the field. As it stands, therefore, our effort may be taken as a sort of exploration in the effort to probe a strongly correlated hadronic system that remains more or less completely uncharted up to date.

Our approach consists of exploiting the presence of half-skyrmions in the skyrmion crystal description of dense baryonic matter

\section{Topology Change}\label{topology}
\index{topology change! dense matter}

In this article, by ``{\it skyrmion}" we will mean baryonic soliton in the general sense, independently of the degrees of freedom involved, i.e., vector mesons, scalar meson etc. in addition to pion. The soliton from the original Skyrme model with pions only,
\be
{\cal L}= \frac{f_\pi^2}{4} {\Tr} \left(\del_\mu U\del^\mu U^\dagger \right)+\frac{1}{32e^2}{\Tr} \left[U^\dagger\del_\mu U, U^\dagger\del^\nu U\right]^2
 \ee
implemented with mass terms, will be denoted $\pi$skyrmion. In this paper, we will be mainly dealing with the {\it skyrmion} and $\pi$skyrmion will be specifically mentioned whenever appropriate.

A key element in the development that follows is the potentially powerful role that fractionized skyrmions, specially half-skyrmions that carry  half baryon numbers, play in dense matter.  There is no established proof that half skyrmions are the most favorable configurations. Indeed in some low-dimension systems (described, e.g., by Karliner and Hen in this volume), other fractionized configurations (e.g., hexagonal) are found to be energetically favored. In string-theory-motivated approaches where instantons in five dimensions figure as baryons as discussed in the third part of this volume by Kaplunovsky, Melnikov and Sonnenschein and also by Sutcliffe, a popcorn structure rather than fractionized solitons can arise.

 In hadron physics in terms of skyrmions in 4 dimensions, only half-skyrmions have been studied. We shall focus on them. How they appear and evolve in baryonic matter as density increases is fully described in the article by Park and Vento in this volume, referred to for short as PV.  We will therefore not go into details that are covered by PV, inviting the readers to their review, also for historical background. The present article recounts the subsequent development that updates -- and revises as needed -- what's given there. What is essential for our discussion is to note that the presence of the half-skyrmion ``phase"\footnote{The term ``phase" used here -- and in what follows, strictly speaking, is a misnomer. There is no order parameter in terms of a local field that characterizes the state involved, so does not belong to the usual Ginzburg-Landau-Wilson paradigm for phase transitions.} in dense matter is generic in the skyrmion description. In fact quite surprisingly the half-skyrmion structure is already present in the $\alpha$ nucleus (with four nucleons) as discussed by Battye, Manton and Sutcliffe in this volume. Furthermore it turns out that its appearance is independent of what other degrees of freedom than that of pion are present. This can be understood by that the Lagrangian we will be using is strictly valid in the large $N_c$ limit, and in the large $N_c$ limit and at high density, baryonic matter is a crystal with the skyrmion fractionized into two half-skyrmions. This is a {\it robust} topological phenomenon, involving the pion field.

The simple way to understand the phase change involved is in terms of the chiral $SO(4)$ coordinates, $(\sigma, \pi_1,\pi_2,\pi_3)$. There is an enhancement of the symmetry~\cite{manton-sutcliffe}
\be
(x^1,x^2,x^3)&\mapsto& (x^1+L,x^2,x^3),\nonumber\\
 (\sigma,\pi_1,\pi_2,\pi_3)&\mapsto& (-\sigma, -\pi_1,\pi_2,\pi_3)
\ee
as the lattice size $L$ is decreased (which corresponds to increasing the density $n$) in the system of skyrmions put in the face-centered-cubic (FCC) crystal. The symmetry enhancement lowers the energy, thereby inducing the phase change. Each cube in this configuration has the baryon number 1/2, i.e., half-skyrmion. On average, $\la\sigma\ra=0$\footnote{The symbol $\sigma$ will be reserved for its connection to the bilinear quark condensate $\la\sigma\ra\propto \la\bar{q}q\ra$. In the literature, it is often used for dilaton. In what follows, the dilaton wlll be denoted as $\phi$.}, which in the QCD variable, quark condensate,  is $\la\bar{q}q\ra=0$. Formally this looks as if chiral symmetry is restored to Wigner phase. We will see, however, that this is not the case. Although the quark condensate is zero averaged over the unit cell, it is locally non-zero giving rise to a chiral density wave. The pion decay constant is non-zero with the hadrons gapped.

\section{Scale-Invariant Hidden Local Symmetry}
\index{scale-invariant hidden local symmetry, sHLS}

While the topology is dictated by the pion, the detail structure of the baryonic matter, i.e., the equation of state (EoS), depends strongly on the degrees of freedom that enter in the dynamics. At high density where short-distance interactions intervene, heavy-mass degrees of freedom are needed. In $\pi$skyrmion, only the Skyrme quartic term figures to capture the short-distance dynamics. We will see that qualitatively new features come in with explicit higher-mass excitations included. Here we include the lowest-lying vector mesons $V\equiv (\rho, \omega)$ and the scalar $\phi$ with the mass $\sim 600$ MeV. The vectors will be incorporated as hidden local fields and the scalar as a dilaton. For this we resort to scale-invariant hidden local symmetry (sHLS for short). We consider  baryons generated from this sHLS Lagrangian as solitons.

\subsection{Vector mesons as hidden local fields}

In the vacuum, that is, matter-free space, the vector-meson mass is big, $m_V \sim  6 m_\pi$, so for low density the explicit account of the vector degrees of freedom is not needed. However in medium with large density, there is a possibility, as we will explain, that the vector meson mass can decrease substantially as density approaches that of chiral restoration and the explicit description of their degrees of freedom is required. A powerful way of introducing the vector degrees of freedom is hidden local symmetry (HLS)~\cite{HLS,HY:PR} that captures this possibility. We will exploit this approach which is closer in spirit to holographic dual models coming from string theory discussed in the last section of this volume.

To bring to light the notion that HLS in low-energy dynamics is natural, it is instructive to see how hidden local fields ``emerge" naturally from a low-energy theory. As will be noted, the same structure can be obtained top-down from string theory.

The idea is simply that the chiral field $U=e^{2i\pi/f_\pi}$ -- which represents the coordinates for the symmetry $SU(N_f)_L\times SU(N_f)_R/SU(N_f)_{L+R}$ -- can be written in terms of the left and right coset-space coordinates as
 \be
U=\xi_L^\dagger\xi_R\label{Up}
 \ee
with transformation under $SU(N_f)_L\times SU(N_f)_R$ as
$\xi_L\rightarrow \xi_L L^\dagger$ and $\xi_R\rightarrow \xi_R R^\dagger$ with
$L(R)\in SU(N_f)_{L(R)}$.  Now the redundancy that is hidden, namely, the invariance under the $local$
transformation
 \be
\xi_{L,R}\rightarrow h(x)\xi_{L,R}
 \ee
where $h(x)\in SU(N_f)_{V=L+R}$, can be elevated to a local gauge invariance~\cite{HY:PR} with the corresponding gauge field
$V_\mu\in SU(N_f)_{V}$ that transforms
 \be
V_\mu\rightarrow h(x)(V_\mu +i\del_\mu)h^\dagger (x).
 \ee
The resulting  HLS Lagrangian \index{hidden local symmetry (HLS)} given in terms of the covariant derivative $D_\mu$ takes the form  (with $V_\mu=g\rho_\mu$):
 \be
{\calL}&=& \frac{F_\pi^2}{4}{\Tr}\left\{|D_\mu\xi_L|^2 + |D_\mu\xi_R|^2
 + \gamma |D_\mu U|^2\right\} \nonumber\\
  & -& \frac{1}{2} \, \mbox{Tr} \left[ \rho_{\mu\nu} \rho^{\mu\nu} \right] +\cdots \label{hls1}
 \ee
where $\rho_{\mu\nu}$ is the $\rho$ field tensor and the ellipsis stands for higher derivative and other higher dimension terms including the gauged Skyrme term. Note that in the power counting with HLS, the vector-meson kinetic energy term is of ${\cal O}(p^2)$. If one parameterizes $\xi_{L,R}=e^{i\sigma/f_\sigma}e^{\mp i\pi/f_\pi}$, gauge-fixing with $\sigma=0$ corresponds to unitary gauge, giving  the usual gauged nonlinear sigma model with a mass term for the gauge field\footnote{The $\sigma$ here, used in the literature for HLS, should not be confused with $\sigma\sim \bar{q}q$ of this article.}. Clearly one can extend such a construction to an infinite tower of HLS vector mesons spread in energy in the fifth dimension. Such a construction has been made and led to the so-called ``dimensionally deconstructed QCD" encapsulated in a 5D Yang-Mills theory~\cite{son-stephanov}. The latter is essentially equivalent in form to the 5D Yang-Mills theory of holographic dual QCD that comes from string theory~\cite{SSmodel}, the main difference lying in the background metric.  As noted by Harada, Matsuzaki and Yamawaki~\cite{HMY}, the Lagrangian (\ref{hls1})  can be thought of as a truncated version of the infinite-tower HLS where {\em all} other than the lowest vector mesons $\rho$ and $\omega$ are integrated out.

The $\omega$ can be put in $U(2)$ symmetry  with the $\rho$, and treated together but we will see that this symmetry breaks down at high density, so we will incorporate $\omega$ as a $U(1)$ local gauge field. For simplicity, however, unless needed otherwise, we write things in $U(2)$ symmetric way.  We will denote the HLS Lagrangians with $U(2)$, $SU(2)$ and $U(1)$ symmetries, respectively, as HLS$_{X}$ with $X=U(2), SU(2), U(1)$ if needed specifically. One can write down a general HLS Lagrangian in power series in covariant derivatives with chiral symmetry breaking terms suitably incorporated.

The  \su2hls Lagrangian, at low-energy scale where the kinetic energy term for the vector mesons is ignorable, is gauge-equivalent to the non-linear sigma model, hence should give more or less the same baryon structure as the $\pi$skyrmion. The expansion in covariant derivatives is a generalization of the usual chiral expansion to one that includes the vector mesons~\cite{HY:PR} (and later the dilaton scalar~\cite{CT}). The feature that distinguishes the HLS approach from others that possess no such symmetry is that HLS Lagrangian predicts what is known as ``vector manifestation (VM)". The VM states that when chiral symmetry is restored with the chiral condensate $\la\bar{q}q\ra\rightarrow 0$, the $\rho$ mass goes to zero as~\cite{HY:PR}
\be
m_\rho\propto g\propto \la\bar{q}q\ra\rightarrow 0.\label{vm}
\ee
This is one of the principal features we will exploit in our approach. We will find that it makes a prediction that has not been discussed in the literature.

The prediction (\ref{vm}) follows from a Wilsonian renormalization group (RG for short) analysis of the HLS Lagrangian matched, via correlators, to QCD at a matching scale $\Lambda_M\sim \Lambda_\chi$ with $\Lambda_\chi$ being the chiral symmetry scale. It is valid in the large $N_c$ limit and, being an RG-based argument, is most likely reliable. But the Lagrangian contains mesons only. Now one might ask, what about baryon degrees of freedom? The answer is that baryons are to arise from the HLS Lagrangian as solitons. The question then is whether the VM structure remains unmodified when skyrmions are considered. As noted above, the skyrmions present at low density fractionize to half-skyrmions at higher density. So the next question is: Does the half-skyrmion structure preserve the VM fixed-point? The answers to these questions, particularly the last one,  remain more or less unanswered. To address the first question, it was established~\cite{kim-vm} that when constituent quarks are introduced hidden local symmetrically, the VM structure is retained provided that the constituent quark becomes massless at the chiral restoration point. This compellingly suggests -- and we will assume -- that the VM fixed point for $\rho$ makes sense also when approached from the half-skyrimion phase.  This assumption will be found to lead to several striking results in the EoS for compact stars as we will show below.

\subsection{Scalar meson as a hidden dilaton}
\index{hidden dilatons}
\index{$f_0(500)$! scalar Nambu-Goldston dilaton}
A scalar meson of low mass $\sim 600$ MeV is an indispensable degree of freedom in nuclear physics. It has figured for decades in phenomenological nucleon-nucleon potentials and in relativistic mean-field approaches for nuclear many body problems. It plays a crucial role not only for nuclei and nuclear matter but also for compact star matter. Although there are model-independent formalisms that establish its existence, now named $f_0(500)$ in the particle data booklet, including its pole position and width, its structure in the light of QCD is still more or less unknown. Its long history, controversies and evolutions are described in a recent review by Pel\'aez~\cite{pelaez}.

In this article, we adopt the notion that the scalar meson needed in nuclear physics is a dilaton, a pseudo-Nambu-Goldstone boson, as proposed by Crewther and Tunstall~\cite{CT}, anchored on the conjecture that there is an IR fixed point in QCD that gives rise to a light-mass scalar, i.e., dilaton. In our previous contribution~\cite{LR-1st} and also described by PV~\cite{PV}, both of which appeared in the first volume, the dilaton also figured.

To clarify how the IR fixed point presents a new approach, let us review what has been done~\cite{LR-1st}. There, recognizing that  the spontaneous breaking of chiral symmetry that generates hadron masses and the explicit breaking of scale invariance by the quantum anomaly in QCD, which brings a length scale, must be connected, the dialton was decomposed into two components,  one ``soft" and the other ``hard," with the soft dilaton $\chi_s$ intervening in the spontaneous breakdown of chiral symmetry and the hard dilaton $\chi_h$ intervening in confinement-deconfinement. By integrating out the latter to focus on the chiral symmetry properties of hadrons, a suitable soft-dilaton-implemented HLS Lagrangian was obtained in \cite{LR09}. Written in unitary gauge for which  $\xi=\xi^\dagger_L=\xi_R =\sqrt{U}$ and with some harmless simplifications, it takes the form (including the pion mass term) for two light flavors (up and down)\footnote{For economy in notation, Eq.~(\ref{lags}) is written with $U(2)$ symmetry for $(\rho,\omega)$. {We are also writing, for simplicity, the Lagrangian in a formally Lorentz-invariant form although in medium the symmetry is broken down spontaneously to $O(3)$. In the applications~\cite{LPR,PKLR} described below the $O(3)$ covariant structure is properly taken into account.} }:
\be
{\cal L}={\cal L}_{\chi_s}+ {\cal L}_{hWZ}\label{lagtot}
\ee
where
\be
{\cal L}_{\chi_s} &=& \frac{F_\pi^2}{4}\kappa^2
\mbox{Tr}\left(\partial_\mu U^\dagger \partial^\mu U\right) + \kappa^3 v^3 \mbox{Tr}{\cal M}\left(U+U^\dagger\right)
\nonumber\\
& &{} - \frac{F_\pi^2}{4} a\kappa^2 \mbox{Tr}\left[\ell_\mu + r_\mu + i(g/2)
( \vec{\tau}\cdot\vec{\rho}_\mu + \omega_\mu)\right]^2 \nonumber\\
& &{} - \textstyle \frac{1}{4} \displaystyle
\vec{\rho}_{\mu\nu} \cdot \vec{\rho}^{\mu\nu}
-\textstyle \frac{1}{4}  \omega_{\mu\nu} \omega^{\mu\nu}
+\frac 12 \del_\mu\chi_s\del^\mu\chi_s + V(\chi_s) \nonumber\\
\label{lags}
\ee
where $\kappa=\chi_{s}/f_{\chi_{s}}$ with $f_{\chi_s}=\la 0|\chi_s|0\ra$, $l_\mu=\del_\mu\xi\xi^\dagger$, and $r_\mu=\del_\mu\xi^\dagger\xi$.
%\be
%B^\mu =  \frac{1}{24\pi^2} \varepsilon^{\mu\nu\alpha\beta}
%\mbox{Tr}(U^\dagger\partial_\nu U U^\dagger\partial_\alpha U
%U^\dagger\partial_\beta U)
%\ee is the baryon current.
For flavor number $N_f < 3$, the well-known 5D topological Wess-Zumino term is absent. However in the presence of vector mesons as in HLS formulation, there are in general three terms  (in the absence of external sources),  called  ``hWZ terms," in the anomalous parity sector that satisfy homogeneous anomaly equation. These terms are of scale-dimension 4, i.e., scale-invariant and hence are not multiplied by $\kappa$.

In the previous works~\cite{LR09}, the dilaton potential was taken to be of the type obtained by summing weak {\it explicit} scale-symmetry-breaking terms to all orders~\cite{grinstein},
\be
V(\chi_{s})=B\chi_{s}^4{\rm ln}\frac{\chi_{s}}{f_{\chi_{s}}e^{1/4}}.\label{potterm}
\ee
 We shall call the scenario based on this dilaton as ``soft-dilaton" scenario. It is important to note that with this potential, the spontaneous breaking of scale symmetry cannot take place without the explicit symmetry breaking. The scheme of \cite{CT} differs from this in that the dilation current $D_\mu=x^\nu \theta_{\mu\nu}$ can be partially conserved in the same way as the axial current is.

Our claim is that the dilaton structure based on the IR fixed point that we shall call ``IR fixed-point scenario" leads to a picture that is simpler, conceptually more elegant and potentially  more predictive for dense matter as it is for elementary particle processes such $K\rightarrow 2\pi$ process. This new approach which supersedes a previous work in a similar line~\cite{dongetal} has recently been applied to massive compact stars~\cite{LPR,PKLR}.

That $f_0(500)$ is a dilaton we shall denote as $\phi$, not as $\sigma$ found in the literature~\cite{CT},  a pseudo-NG boson on the same footing as the octet psudoscalar pseudo-NG bosons, is anchored on the (presumed)  existence of an IR fixed point.  We will return to this matter in the next subsection.
In what follows, we shall simply adopt this approach. Our main reason is that it has the power to allow the scalar dilaton to be treated as a {\it local field}, an appealing theoretical justification for the long-standing practice in nuclear physics. In fact, the large width associated with $f_0(500)$ appears to be accountable in a systematic power counting in the scale-chiral perturbation theory, that we refer to as sChPT or $\chi$PT$_\sigma$~\cite{CT}. This clearly does away with the tortuous summation of high-order chiral perturbation series in the standard 3-flavor chiral perturbation theory or equivalently 3-flavor HLS theory.

Since the dilaton field that we shall denote as $\phi$, $\phi=f_\chi \ln \frac{\chi}{f_\chi}$, can be put on the same footing as the octet pseudoscalar pseudo-NG bosons, it is straightforward to implement it  to  the HLS Lagrangian.  Starting with the  Lagrangian written in power series in chiral order, one can incorporate the dilaton $\phi$ in consistency with the scale-chiral counting.  This is done by assigning the deviation of the QCD gauge coupling $\alpha_s=\frac{g_s^2}{4\pi}$ from the IR fixed point, $\Delta \alpha_s=\alpha_s-\alpha_{IR}$, the chiral order ${\cal O} (M)$ (where $M$ is the quark mass  matrix)
\be
{\O} (\Delta\alpha_s) \sim {\O} (p^2) \sim {\O}(M).\label{counting}
\ee
Although we are dealing with the flavor $SU(3)$ for the dilaton, we will, for applications, be focusing on 2-flavor systems, so we will be projecting out the $SU(2)$ sector. Then it is a good approximation to ignore the effect of the anomalous dimension $\beta^\prime$ of the stress tensor of the trace anomaly in the matter Lagrangian,  which is ${\O}(M)$ or higher order in the matter sector~\footnote{This corresponds to setting $c_1=c_2=1$ in Ref.~\citen{CT}}. Then the resulting sHLS Lagrangian can be written in the same form as Eq.~(\ref{lags}) with $\chi_s$ replaced by $\chi$ except that the dilaton potential now is given by
\be
V(\chi)=a\left(\frac{\chi}{f_\chi}\right)^4+b\left(\frac{\chi}{f_\chi}\right)^{4+\beta^\prime}\label{potCT}
\ee
where $a$ and $b$ are constants to be determined. The trace of the energy-momentum tensor including the quark mass term  is given by
\be
\theta_\mu^\mu=b\beta^\prime \left(\frac{\chi}{f_\chi}\right)^{4+\beta^\prime} -\left(\frac{\chi}{f_\chi}\right)^3{\Tr} \left({\cal M}U^\dagger+U{\cal M}^\dagger\right).
\ee

\subsection{Scale symmetry in dense matter as an emergent symmetry}
\index{emergent symmetry! scale invariance in matter}

There is, up to date, no definitive confirmation, theoretical or experimental, for the presence of an IR fixed in QCD. Neither has it been proven to be absent. In fact there is at least one support for it: a numerical stochastic perturbation calculation~\cite{stochastic} for two flavors that ``votes" for its existence.  However it would have to be supported by a lattice simulation for $N_F=3$. In the Crewther-Tunstall scheme which justifies the scalar $f_0(500)$ to be described by a local field, the scale symmetry can be spontaneously broken and the chiral condensate $\la\sigma\ra$ can be nonzero at the IR fixed point~\cite{CT}. Whether or not such a scenario is viable remains to be seen.

That the scalar in the relativistic mean field model effective in nuclear matter must be a chiral scalar, but transmutes to  the fourth component of the chiral four-vector as the chiral restoration is approached with the triplet pion and the scalar joining the chiral multiplet~\cite{sasakietal} can be understood as suggesting that the scale symmetry, hidden in the matter-free vacuum, can show up as an ``emergent symmetry" in medium even if IR fixed point is absent in the matter-free space and gets locked to chiral symmetry in dense medium\footnote{This possibility was suggested to us by Koichi Yamawaki by private communications. An illuminating observation made by Yamawaki and his colleagues~\cite{dilatonicHiggs} is that the familiar linear sigma model contains hidden scale symmetry that gets manifested when a parameter that governs the potential, denoted $\lambda$, is fine-tuned. By dialling from $\lambda=\infty$ to $\lambda=0$, it is seen that the linear sigma model that encodes the Standard Model captures the physics that ranges from that of the nonlinear sigma model with no light scalars to a scale-invariant theory with a massless dilaton. This is the scenario discussed in Ref.~\citen{sasakietal} for baryonic matter going from normal nuclear matter to highly dense matter. }.  This can be seen as follows. When the topology change  takes place, as mentioned in Section \ref{topology},  from skyrmions to half-skyrmions at $n_{1/2}$, the chiral condensate $\sigma\propto \la\bar{q}q\ra$ vanishes globally, so the QCD gauge coupling $\alpha_s$ will be running solely due to the trace anomaly with the QCD scaling $\Lambda_{QCD}$ -- which depends on density in medium. The density then provides an IR scale in this system. It is intriguing that the scale symmetry that emerges here can be considered as a hidden scale symmetry along the line discussed in connection with dilatonic Higgs in particle physics~\cite{dilatonicHiggs}. It seems possible that the scale symmetry will manifest itself from the half-skyrmion phase with a vanishing dilaton mass in a manner similar to the manifestation of hidden local symmetry in the vicinity of the VM fixed point.  It is noteworthy in the Crewther-Tunstall scenario that in the chiral limit with the vanishing quark mass, the dilaton current is partially conserved, an analog to PCAC. The exact conservation is arrived at the IR fixed point.

\section{Skyrmion Matter}
\index{skyrmion matter}
While a skyrmion matter with sHLS Lagrangian with the dilaton potential (\ref{potterm}) has been studied~\cite{maetal-shls}, the analysis with the potential (\ref{potCT}) is not yet available at the time of writing this article. It is however feasible to map robust features extracted from the sHLS skyrmion matter, largely free from the detail structure of the potential, to the ``bare parameters" of the Lagrangian and then analyze the property of dense matter via RG using the technique of continuum Lagrangian. This will be described in the next section. Here we discuss what we consider to be generic structures encoded in the skyrmion description, more or less independently of the dilaton degree of freedom.

As mentioned above and clearly described by PV~\cite{PV}, one of the most robust features in the dense skyrmion description is the topology change from skyrmions to half-skyrmions at some density $n_{1/2}$. The location of $n_{1/2}$ depends on what degrees of freedom are included in the Lagrangian but the presence of the changeover is independent of them. It will turn out that the most plausible value for $n_{1/2}$ in confronting Nature is $\sim 2n_0$. For the moment, we do not need its precise value, only that it is above normal nuclear matter density. We shall now discuss a few striking features predicted by the skyrmion structure of dense matter.

\subsection{The symmetry energy}
\index{symmetry energy}

The most non-trivial and potentially powerful prediction is the cusp structure in the symmetry energy at $n_{1/2}$ discovered in Ref.~\citen{cusp}.  Consider asymmetric nuclear matter (with neutron excess). The energy per particle of asymmetric nuclear matter is given by
\be
E(n,\delta)=E_0(n)+E_{sym}(n) \delta^2 +\cdots
\ee
 where $\delta=(N-P)/(N+P)$ with $N(P)$ the number of neutrons (protons) and the ellipsis stands for higher orders in $\delta$.  We are interested in calculating the symmetry energy $E_{sym}$ using the sHLS Lagrangian. The symmetry energy is given by a $1/N_c$ term in the skyrmion matter energy, so it requires to be collective-quantized~\cite{cusp}. Unfortunately it is not known how to collective-quantize $A=\infty$-skyrmion crystal (for $\delta\neq 1$ infinite matter) with the sHLS Lagrangian.  However to compute the symmetry energy, we may take a pure neutron matter~\cite{klebanov}.  Since we are dealing with topological structure, we may also take the Skyrme Lagrangian with the vector mesons considered to be integrated out\footnote{The vector degrees of freedom are unlikely to qualitatively modify the result. Their effects are being investigated at the time of this writing~\cite{ma}.}. In Ref.~\citen{cusp},  the dilaton was retained. Collective-rotating the skyrmion neutron matter  with a single set of collective coordinates $U(\vec{r}, t) = A(t) U_0(\vec{r}) A^\dagger (t)$ where $U_0(\vec{r})$ is the static crystal configuration with the  lowest energy for a given density, the canonical quantization leads to
\be
E^{\rm{tot}} = A M_{\rm{cl}}
+ \frac{1}{2A \lambda_{I}} I^{\rm{tot}} (I^{\rm{tot}}+1),
\ee
where $M_{\rm{cl}}$ and $\lambda_{I}$ are, respectively, the mass and the moment of inertia of the single cell. Both are given as integrals over the crystal configuration $U_0$ and dilaton configuration. $I^{\rm{tot}}$ is the total isospin which would be the same as the third component of the isospin $I_3$ for pure neutron matter. This suggests taking,  for $\delta\equiv (N-P)/(N+P)\lsim 1$,
\be
I^{\rm{tot}}=\frac 12 A\delta.
\ee
Thus the energy per nucleon in an infinite matter  is
\be
E=E_0 +\frac{1}{8\lambda_I}\delta^2.\label{E}
\ee
with $E_0=M_{\rm cl}$. This leads to the symmetry energy
\be
E_{sym}=\frac{1}{8\lambda_I}.
\ee

\begin{figure}[ht]
\begin{center}
\includegraphics[scale=0.7]{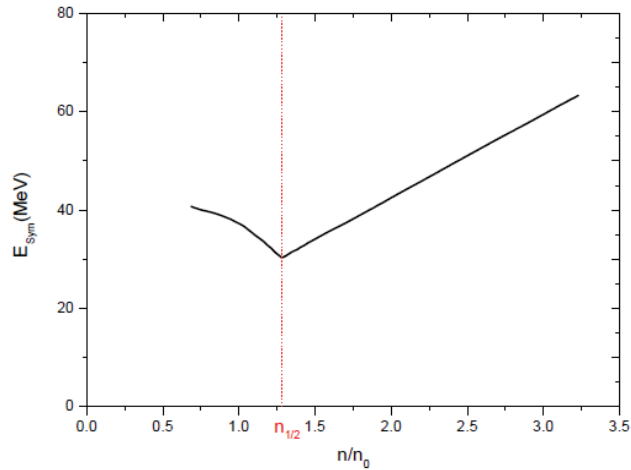}
\caption{A schematic form of the symmetry energy given by the collective rotation of the skyrmion matter in the Skyrme model implemented with a dilaton field. The parameters chosen for the calculation in \cite{cusp} are $f_\pi=93$ MeV, $1/e^2\approx 0.03$ and  the dilaton mass $\sim 600$ MeV. The cusp is located at $n_{1/2}$. The values for the symmetry energy and the location of the cusp $n_{1/2}$ depend on parameters and are given for illustrative purpose.}
\label{cusp}
\end{center}
\end{figure}

In Fig.~\ref{cusp} is shown the symmetry energy with a cusp structure predicted by the topology change at $n_{1/2}$. The quantitative structure, such as the location of $n_{1/2}$, the energy scale etc., will depend on the parameters of the model. What is significant is its robust {\it qualitative} nature. It is only mildly sensitive to the mass of the dilaton, so we expect that the topological structure will not be sensitive to the character of the dilaton potential -- (\ref{potterm}) or \ref{potCT}) --  either.

With the rich array of experimental data available, the symmetry energy is fairly well determined up to nuclear matter density $n_0$. At least up to $n_0$, Nature shows no clear indication for such a cusp structure of Fig.~\ref{cusp} seen in the skyrmion crystal: While $E_{sym}$ is unknown above $n_0$, the decrease toward $n_0$ is not visible in experiments. So one might question the viability of the predicted feature. It turns out however, as will be shown in the next Section, that the cusp, unrealistic though it might appear, is actually consistent with what is given by nuclear effective theory at the leading order in many-body correlations with the sHLS Lagrangian (with baryons introduced explicitly as described below). Our main thesis is that this topological structure gives instead a novelty to the nuclear tensor forces that govern the symmetry energy. In fact it will turn out to give also a crucial clue to the structure of EoS at high density.

\subsection{Nucleon mass and parity doubling}

Another unexpected outcome from the transition from skyrmions to half-skyrmions is the behavior of the baryon mass as the density goes above $n_{1/2}$. The mass of a single baryon estimated from the skyrmion crystal decreases smoothly as density approaches $n_{1/2}$ from below but at $n_{1/2}$ the quark condensate, while nonzero locally, vanishes globally and the baryon mass stays constant and non-zero. The pion decay {constant} also remains non-zero. Since hadron masses are gapped, the chiral symmetry remains broken.

\begin{figure}[ht]
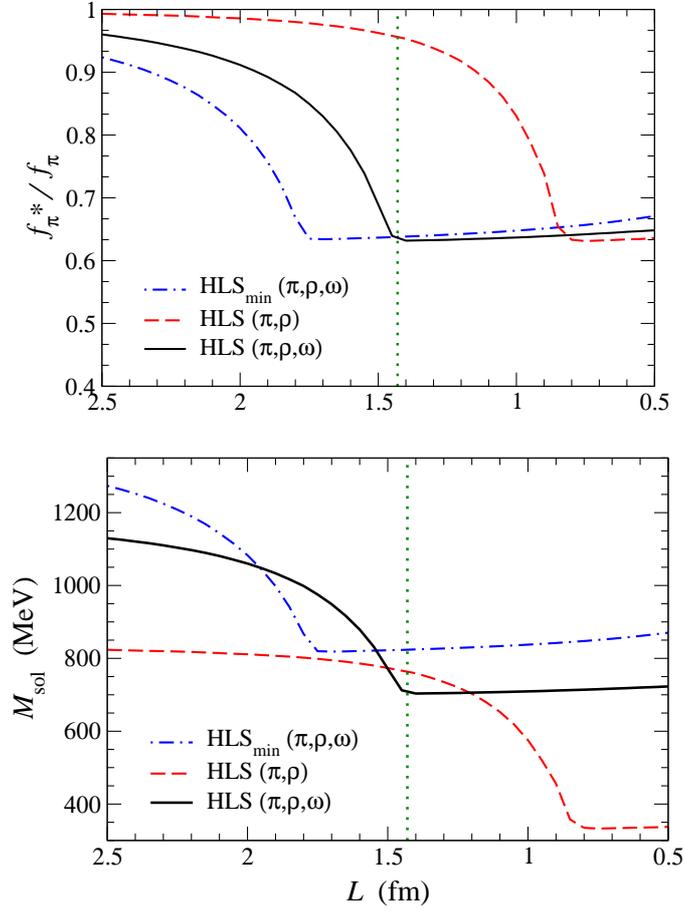

\begin{center}
\includegraphics[height=6cm]{HLMR-fpistar}
\includegraphics[height=6cm]{HLMR-mnstar}
\caption{$f_\pi^\star$ (upper panel) and $M_{\rm sol}^\star$ (lower panel) for decreasing crystal size (increasing density) taken from Ref.~\citen{maetal-hls}. }
\label{pseudogap}
\end{center}
\end{figure}

As of today, there is no result available in the IR-fixed-point scenario. However the results for sHLS with the soft-glue scenario and HLS (without dilaton) show roughly the same structure. We therefore give the results of HLS theory without dilaton~\cite{maetal-hls}. We do not expect that there will be a basic difference between the two scenarios since the quantities calculated are not expected to be sensitive to how the scale symmetry manifests itself in the range of density considered. The results are given in Fig.~\ref{pseudogap}.

Some explanations on Fig.~\ref{pseudogap} are in order here. For an illustrative purpose, we take three variations on the degrees of freedom entering in the hidden local sector. (a) The full, what we consider to be the only reliable, case is the HLS$(\pi,\rho,\omega)$. It embodies complete local hidden gauge invariance with $U(2)$ symmetry for the vector mesons and  contains all three  hWZ terms encoding anomaly, all expanded to ${\cal O} (p^4)$ chiral order. (b) The HLS$_{min}$$(\pi,\rho,\omega)$ is the ${\cal O}(p^2)$ part of HLS$(\pi,\rho,\omega)$ plus one term reduced from the three hWZ terms, assuming vector dominance in photon induced processes and taking the $\rho$ mass to be ``heavy" in its equation of motion. The first assumption is harmless but the second is not consistent with the VM of HLS, which requires that the $\rho$ be treated as light as the $\pi$.   (c) The third, HLS$(\pi,\rho)$, is without the hWZ terms, hence the $\omega$ is decoupled from the system.

Here are some notable, and potentially significant, observations.
\begin{enumerate}
\item One sees that the $\omega$ plays an extremely important role, as observed in a different context~\cite{paengetal-rg}, in the HLS-skyrmion approach to baryonic matter. It affects qualitatively the location of $n_{1/2}$. This also highlights the crucial role of {\it all three} of the hWZ terms. It is important that in all three cases, the topology change does take place although the details differ depending on the degrees of freedom.
\item In all three cases, both the  baryon mass $m_N^\star$ (which is equal to the soliton mass in the large $N_c$ limit) and the pion decay constant $f_\pi^\star$ drop smoothly up to $n_{1/2}$ and then stabilize to a density-independent constant at higher densities. This reflects that both quantities are composed of a part that goes to zero as $\la\bar{q}q\ra\rightarrow 0$  and a part that remains more or less independent of density.
\item It comes out to a very good approximation in the case (a) that $m_N^\star/m_N\approx f_\pi^\star/f_\pi$  throughout the range of density explored. One could understand this simply as an indication that the large $N_c$ approximation holds. In the large $N_c$ limit, $m\sim \zeta f_\pi$ with $\zeta$ a scale-independent ${\cal O}(\sqrt{N_c})$ constant.
\item In the case (a), one can write
\be
m_N^\star/m_N\approx A+\Delta (\la\bar{q}q\ra)
\ee
with $A\sim (0.7-0.9)$ and $\Delta (0)=0$. This means that when the quark condensate averages to zero for $n>n_{1/2}$, a large portion of the baryon mass remains nonzero. Although the bilinear quark condensate vanishes on average, it is non-zero locally and has a chiral-density-wave structure~\cite{ma-cdw}. There is parity-doubling although pions are still present. This picture is also arrived at using a renormalization-group technique used in Ref:~\citen{paengetal-rg} with the continuum Lagrangian. This suggests persuasively the robustness of the structure. At high temperatures, there are lattice indications for such an ``apparent chiral-invariant" mass~\cite{temperature-m0}.
\item The half-skrymion phase resembles what's known in the literature as ``quarkionic phase"~\cite{quarkyonic}. We consider this skyrmion crystal structure to be a microscopic description of that phase in hadronic variables. In fact, the EoS described in Ref.~\cite{PKLR} captures the essential features of the quarkyonic star of Ref.~\cite{quarkyonicstar} without resorting to quark variables. This may be formulated as a case of the Cheshire Cat phenomenon described by Nielsen and Zahed in this volume.

\end{enumerate}

\section{Mapping 1/2-Skyrmions to ``Bare" sHLS Lagrangian}

It is not feasible at present to work out realistically the skyrmion structure for complex nuclei and dense matter. It is an extremely daunting mathematical problem. There is some progress in light nuclei with $\pi$skyrmion as discussed in this volume but it will require a breakthrough to be able to address quantitatively and with confidence compact-star matter with sHLS Lagrangian starting from the
crystal structure. Here we would like to bypass the great difficulty by mapping what is considered to be the robust features provided by the skyrmion crystal structure discussed above, i.e., the topology change,  to the ``bare" parameters of sHLS Lagrangian and then apply renormalization-group (RG) strategy to many-body problems involved in strongly-coupled dense medium. For this we will follow the recent developments~\cite{LPR,PKLR}.

\subsection{Matching to QCD correlators}

In order to map what we learn from the skyrmion structure to the sHLS continuum Lagrangian, we consider matching the effective Lagrangian, e.g., (\ref{lagtot}) with (\ref{lags}) given by the IR-fixed-point structure, to QCD via various correlators, i.e., vector axial-vector, tensor etc. correlators, at the matching scale $\Lambda_M$.  The matching endows the parameters of the Lagrangian with ``intrinsic density dependence (IDD)" of QCD via the dilaton condensate $\la\chi\ra$, the quark condensate  $\la\bar{q}q\ra$,  the gluon condensate $\la (G_{\mu\nu})^2\ra$, etc., all of which slide with the ``vacuum" modified by density.  Given such an sHLS, to do many-body calculation, one way is to resort to what is called  ``double decimation RG" calculation. The first is to obtain the $V_{lowk}$ and the second is, using the  $V_{lowk}$ potential, to do a sophisticated many-body calculation to arrive at, and then fluctuate around, the Landau Fermi-liquid fixed point. This is detailed in Ref.~\cite{PKLR} following the well-formulated approach~\cite{doubledecimation}.

The  topology change at $n_{1/2}$ separates the density regime into two regions, Region-I for $n<n_{1/2}$ and Region-II for $n\geq n_{1/2}$. Combining experimental information available in R-I and what is inferred from presently available observables from compact stars for R-II, the IDD bare parameters of the Lagrangian can be written concisely in terms of three scaling parameters $\Phi$, $\kappa_V$ for $V=(\rho,\omega)$,
\be
\Phi (n)=\frac{f_\chi^\star}{f_\chi}, \ \kappa_V (n)=\frac{g_V^\star}{g_V},
\ee
where $f_\chi$ is the dilaton decay constant related to the dilaton condensate $f_\chi=\la \chi\ra$. What makes the IR-fixed point approach predictive in dense matter physics is that as the vacuum is changed by the increasing density, the spontaneous breaking of scale symmetry characterized by the dilaton condensate is ``locked" intricately to  the spontaneous breaking of chiral symmetry. In the soft-glue scenario, this locking was assumed. Here it is automatic.

The analysis  led to the following IDDs of the parameters~\cite{PKLR}:
\be
\frac{m_N^\star}{m_N}\approx \frac{m_\chi^\star}{m_\chi}\approx \kappa_V^{-1}\frac{m_V^\star}{m_V}\approx \Phi (n)
\ee
and
\be
\frac{m_\pi^\star}{m_\pi}\approx \sqrt{\Phi(n)}.
\ee
By the locking of chiral symmetry to scale symmetry, the pion decay constant equals the dialton decay constant $f_\chi\approx f_\pi$, so
\be
\frac{f_\pi^\star}{f_\pi}\approx \Phi(n).\label{piscaling}
\ee
The key point for what follows is that apart from the vector mesons, the IDD in hadron masses and coupling constants is dictated by the dilaton condensate.  It should be noted that the IDD of vector mesons differs by a factor of $\kappa$ from the others. This is due to HLS of the vector mesons which will play an important role in the structure of the nuclear tensor forces, essential for the EoS of compact stars.

The nuclear dynamics involved in going from dilute to dense medium in the RG framework we are adopting is controlled by how the IDD parameters slide as density is increased.  The essential features in R-I and R-II of their role are as follows~\cite{PKLR}.
\begin{enumerate}
\item The topology change induces for $n\geq n_{1/2}$ a drastic change in the behavior of certain bulk properties of nuclear matter, notably the symmetry energy mentioned above and to which we will return below. This change takes place most crucially in the nuclear tensor forces. As we will show, this feature can be exploited to translate the effect of topology change to the behavior of the ``bare" parameters of the baryonic Lagrangian obtained by incorporating baryon fields to sHLS Lagrangian that we will refer to as bsHLS.
\item In R-I, both $\Phi$ and $\kappa_V$ are fairly well determined. Up to the equilibrium nuclear matter density $n_0$, the scale-chiral symmetry locking $f_\chi\approx f_\pi$ provides information on the scaling $\Phi$. This is because the property of $f_\pi$ can be reliably deduced from experiments, namely, in pionic atoms. It seems reasonable that this property can be extended to somewhat above $n_0$, say, to $n_{1/2}$. As for $\kappa_V$, we find it fairly independent of density, so we will simply take it to be $\kappa_V=1$ in R-I.
\item The IDD-scaling property of the R-II region, due to the paucity of both experimental data and trustful theoretical tools, is much less clear. However if one accepts the topology change at $n_{1/2}$ suggested by the skyrmion crystal model, one can exploit the information on the EoS provided by massive compact stars to deduce the scaling behavior of both $\Phi$ and $\kappa$ in R-II.  It was found~\cite{PKLR} that the nucleon mass stops dropping at $n_{1/2}$ and goes to a constant $A\approx (0.7-0.8)$ as in the skyrmion crystal description mentioned above.  In terms of large $N_c$, $m_N\propto f_\chi$, so this scaling applies to the scaling function $\Phi$. This is because it is the dilaton condensate that takes over in R-II.

In contrast, the property of $\kappa_\rho$ is dictated by the vector manifestation of the $\rho$ meson, namely, that the $\rho$ mass behaves proportionally to the hidden gauge coupling $g_\rho$ which goes to zero as the density approaches the vector manifestation fixed-point density $n_{VM}$ which is located near but slightly below -- not precisely on -- the chiral restoration density $n_c$~\cite{HY:PR}. It is to go to zero, near the vector manifestation density $n_{VM}$, linearly in $(n_{VM} -n)$.\footnote{If the inhomogeneous chiral density wave were absent as in the consideration of \cite{HY:PR}, then the $\rho$ mass would go proportionally to the bilinear quark condensate $\la\bar{q}q\ra$ that vanishes as the VM fixed point is approached. In baryonic medium, the approach to the VM fixed point will not be linear in the bilinear quark condensate. In fact in the skyrmion crystal picture, the bilinear condensate, while locally nonzero, vanishes on average for $n>n_{1/2}$ while the order parameter for chiral symmetry, which may be in the form of multiquark condensate, remains non-zero. This is similar to what's described as ``quarkyonic phase." }  {This feature, which is forced on the EoS at $n>n_{1/2}$ by massive compact-star observations, closely resembles, and is possibly connected in a fundamental way to, what happens in sHLS theory applied to techni-$\rho$ in going beyond the Standard Model. There the mass of the $\rho$ meson is argued to be scale-invariant~\cite{dilatonicHiggs}.} On the other hand, the breakdown of $U(2) $ symmetry for the vector mesons, as strongly indicated by the EoS at $n>n_{1/2}$,  makes $\kappa_\omega$ deviate strongly from the behavior of $\kappa_\rho$. This feature, while most likely crucial in compact-star matter, is not of concern for this paper, so will not be further addressed.
\end{enumerate}

\subsection{Intrinsic density dependence (IDD) and nuclear tensor forces}
\index{IDD, intrinsic density dependence! in skyrmion matter}

We now show that the cusp in the symmetry energy Fig.~\ref{cusp} can be reproduced by a many-body calculation with the bsHLS Lagrangian endowed with the IDD's given above.

In the $V_{lowk}$ approach to the symmetry energy, it is the tensor force component of $V_{lowk}$ that plays the dominant role. The tensor force constructed with bsHLS  consists of two terms, one one-$\pi$ exchange and the other one-$\rho$ exchange. Call them $V_\pi^T$ and $V_\rho^T$ respectively. There are two important features with these forces. One, which is well-known, is that the pion tensor and the $\rho$ tensor, while having the same radial function with different masses, come with opposite signs and different overall constants $R^2_{\pi,\rho}$ multiplying the radial functions. And the other, which is not widely known, is the remarkable observation, at present numerical, awaiting a rigorous proof, that the tensor force does not get renormalized under the renormalization-group flow,  that is, the RG beta function for the tensor force is zero, not only in the matter-free vacuum but also in medium, i.e., in the second decimation~\cite{KLPR}.

\begin{figure}[ht]
\begin{center}
\includegraphics[width=9.cm]{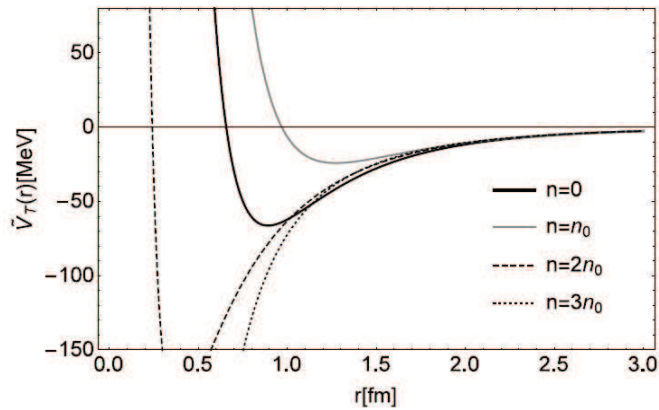}
\caption{Sum of $\pi$ and $\rho$ tensor forces in units of MeV vs. density $n/n_0$. Taken from \cite{PKLR}.}
\label{tensor-pklr}
\end{center}
\end{figure}

In medium, the pion tensor remains unchanged by density over the range of density involved. This has been numerically verified. This results from  what one might interpret as  protection by nearly perfect chiral symmetry. In contrast, the $\rho$ tensor force increases in magnitude in R-I as the $\rho$ mass drops proportionally to $\Phi$ with a constant overall coefficient in front, $R^2_\rho\approx 1$,  so the net tensor force decreases because of the increasing cancellation between the two forces as density approaches toward $n_{1/2}$. However above $n_{1/2}$, the hidden gauge coupling constant drops subject to the VM making the coefficient multiplying the radial part $R^2_\rho$  fall rapidly, scaling as $R^2_\rho\approx \Phi^4$, and thereby drastically quenches the $\rho$ tensor force. Consequently as density exceeds $n_{1/2}$, the $\pi$ tensor completely takes over and further increases the net tensor force strength.  This phenomenon is explained in detail in Ref.~\citen{PKLR,LR-tensor}.  The upshot of what transpires  is illustrated in Fig.~\ref{tensor-pklr} taken from Ref.~\citen{PKLR}.

\begin{figure}[ht]
\begin{center}
\vskip 0.6cm
\includegraphics[width=7.5cm]{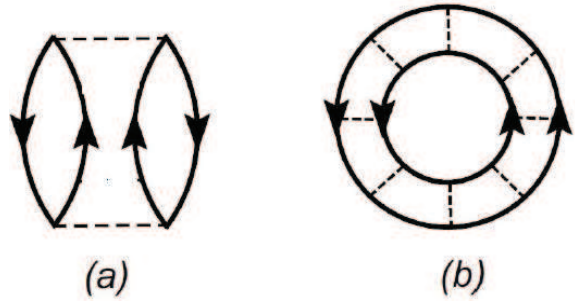}
\caption{Goldstone diagrams contributing to the symmetry energy, the summing of which corresponds to doing the second decimation in the $V_{lowk}$ scheme.
The dotted line represents the sum of $\pi$ and $\rho$ exchanges with the IDD included in the parameters as described in the text. The figure (a) is the leading order and (b) high-order ring diagram series.}
\label{goldstone}
\end{center}
\end{figure}

We now explain how the qualitative feature of the tensor force structure as seen in Fig.~\ref{tensor-pklr} gives the cusp structure observed in Fig.~\ref{cusp}. For this, we recall that in the first decimation $V_{lowk}$ does not undergo renormalization.  Hence we can, with confidence, take the tensor force  given by  Fig.~\ref{tensor-pklr} in performing the second decimation. In field theoretic many-body approaches, this means calculating Goldstone diagrams of the kind given in Fig.~\ref{goldstone}.  Because of the tensor structure, the Hartree term does not contribute, so Fig.~\ref{goldstone}(a) gives the leading contribution.  The second figure, Fig.~\ref{goldstone}(b), represents an all-order graph in the ring-diagram approximation. This technique summing the ring diagrams has met with a great success both in finite nuclei and nuclear matter as one can see in the recent review~\cite{ringdiagram}. For our purpose, we limit to the leading-order term. We will come back to the ring-diagram calculation for a confrontation with Nature.

To proceed, we exploit the well-known fact that the nuclear symmetry energy is dominated by the tensor force. Thus with the iterated net tensor force, Fig.~\ref{goldstone}(a) is to give the symmetry energy in the leading order.  Since the intermediate state excited by the tensor force peaks strongly at energy of $\sim 200-300$ MeV, we can use the closure approximation
\be
E_{sym}\approx c\frac{\la (V^T)^2\ra}{\bar{E}}
\ee
with $c$ a dimensionless constant and $\bar{E}\approx (200-300)$ MeV. This then immediately leads to the result that exactly reproduces the skyrmion calculation: The symmetry energy drops from below and up to $n_{1/2}$ and then turns up and increases roughly linearly in density above $n_{1/2}$.   Thus with the topology change, the cusp in $E_{sym}$ at $n_{1/2}$ is inevitable. In contrast, without the topology change, $E_{sym}$ will continuously decrease to zero at $\sim 3n_0$, with an important consequence on the symmetry energy  as is discussed below.

\begin{figure}[hb]
\begin{center}
\includegraphics[width=12cm]{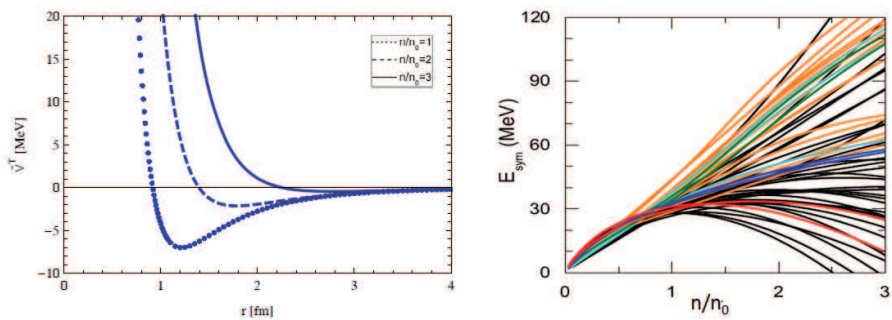}
\caption{Sum of $\pi$ and $\rho$ tensor forces in units of MeV vs. density $n/n_0$ without  topology change at $n_{1/2}$ (left panel) and the ``symmetry energy jungle" given by various nuclear models that {\it reproduce correctly} the saturation properties at $n_0$, reproduced from Ref.~\citen{chen}.}
\label{Esym-wilderness}
\end{center}
\end{figure}
Suppose that there were no topology change at $n_{1/2}$. Then the continuously increasing $\rho$ tensor would cancel away the attraction of the $\pi$ exchange, with the net tensor force going to zero at $n\sim 3n_0$ as in Fig.~\ref{Esym-wilderness} (left panel). The symmetry energy would then look like the curve that vanishes at $\sim 3n_0$, known as ``supersoft" symmetry,  in the ``symmetry energy wilderness" depicted in Fig.~\ref{Esym-wilderness} (right panel), given by phenomenological nuclear models in the market (taken from Ref.~\citen{chen}). Although it may appear to be at odds with the presently accepted gravity theories, such a supersoft symmetry energy has no yet been excluded by rigorous theoretical considerations. However our bsHLS description, if correct, will rule it out.

It has been argued before~\cite{PKLR} that one can take the agreement between the skyrmion at order ${\cal O}(1/N_c)$ and the closure approximation at the leading order in $V_{lowk}$ RG as a support to the procedure of mapping what is considered as robust skyrmion properties to the ``bare" parameters of sHLS Lagrangian. This provides a rationale for the continuum Lagrangian analyses given below for compact-star properties.

While the apparent non-renormalization  may be indicative of the tensor force being at the fixed point, it does not of course imply that the symmetry energy itself will not be modified by higher order correlations, such as Fig.~\ref{goldstone}b, that enter in the second decimation. We will see below that the symmetry energy does in fact get renormalized by higher orders and the cusp gets smoothed while leaving a tell-tale signal of the soft-to-hard change in the EoS. This is because, although dominated by the tensor force, the symmetry energy does also receive non-ignorable contributions from other components of the force. Unsurprisingly, the symmetry energy itself  is not a fixed-point quantity. This is unlike the monopole matrix element in the shell evolution in exotic nuclei studied by Otsuka et al.~\cite{otsuka} which cleanly picks out the tensor force component that points to a fixed-point quantity.

\section{Applications to Compact Stars}

While the skyrmion description either in crystal framework or otherwise cannot at present access directly massive compact stars, the double decimation RG approach using the $V_{lowk}$ using the HLS Lagrangian implemented with the dilaton and baryons (bsHLS) endowed with IDDs  can and has been applied to make certain predictions, giving a novel structure to the EoS of dense compact stars. The details are given in Ref.~\citen{PKLR}, so we will be brief in reviewing the main results. We will see how it works first in R-I, where experiments provide mostly accurate information up to density $n_0$ and then see how it works beyond the experimentally known regime for $n>n_{1/2}$.

\subsection{Phenomena in R-I}

As mentioned, the IDDs of the Lagrangian in R-I are dictated by the dilaton condensate $f_\chi$. All the masses of the degrees of freedom involved, denoted generically $M$, other than the pion mass scale in density with $\Phi$ as
\be
M^\star/M\approx \Phi\equiv f_\chi^\star/f_\chi.
\ee
The pion mass scales differently because the chiral symmetry breaking term with quark mass has the scale dimension 1 and the pion decay constant gets locked to the dilaton condensate $f_\chi$ as $f_\pi\approx f_\chi$. Therefore the pion mass turns out to scale in density as
\be
m_\pi^\star/m_\pi\approx\sqrt{\Phi}.
\ee
Fortunately for numerical analyses,  the density dependence of $\Phi$ is known at least up to $n_0$ because $f_\pi^\star$ is measured up to near the equilibrium density and the standard chiral perturbation theory can be applied in between. What remains to be  determined is the scaling in density of the coupling constants. The relevant quantities are found to be
\be
\frac{g^\star_{\pi NN}}{g_{\pi NN}}\approx \frac{g_A^\star}{g_A}\approx \Phi, \ \ \frac{g_{V NN}^\star}{g_{V NN}}\approx   \frac{g_{s NN}^\star}{g_{s NN}}\approx 1.
\ee

With the known vacuum values of the parameters, the above scalings completely determine the bsHLS Lagrangian with which one can do the $V_{lowk}$ calculations.

\subsubsection{The C14 dating beta decay}

\begin{figure}[ht]
\vskip 0.6cm
\begin{center}
\includegraphics[width=9.cm]{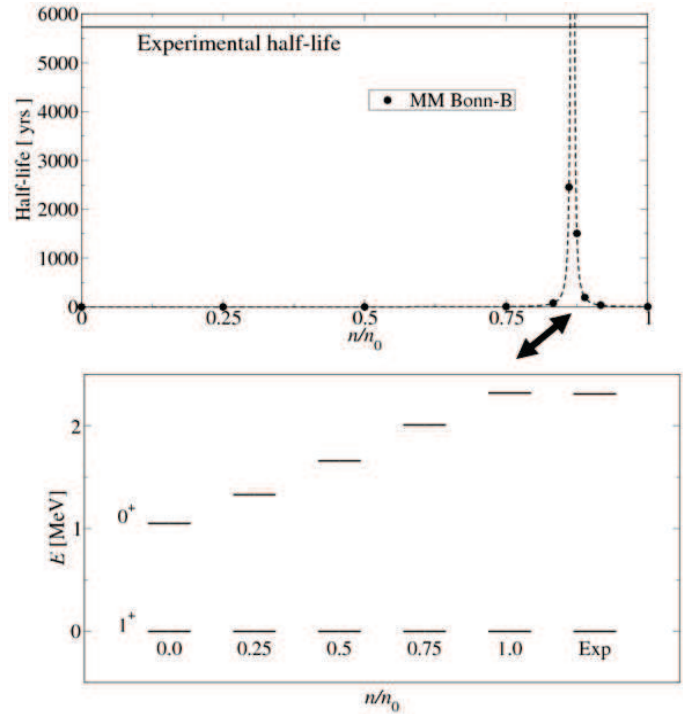}
\caption{The half-life of $^{14}$C as a function of nuclear density (upper panel) and the splitting  between the $1_1^+$ and $0_1^+$ levels in $^{14}$N vs. density (lower panel). Taken from Ref.~\citen{holt-c14}. }
\label{c14}
\end{center}
\end{figure}

What may be taken as a beautiful confirmation of the mapping strategy applicable in R-I is the C14 dating. The long half-life, 5730 years, of C14 can be neatly explained in terms of the IDD in the structure of the tensor forces~\cite{holt-c14}. The density dependence in the ``bare" potential, notably in the quenching of the tensor force at increasing density discussed above, brings out a delicate cancellation in the Gamow-Teller transition matrix element. The p-shell nucleon involved in the transition feels density in the range $(0.75-1)n_0$ and the cancellation is ``fine-tuned" by Nature to that range of density. This is seen in the upper panel of Fig.~\ref{c14}. That the cancellation is not a pure accident but a robust effect of the quenched tensor strength can be seen in the lower panel where the energy of the lowest excited $0^+$ state in $^{14}$N comes out correctly at the same density range. The former deals with the axial current and the latter with the Hamiltonian and there is no simple and direct relation between the axial current and the Hamiltonian~\footnote{For the vector current, e.g., isovector magnetic moment, there is the Siegert theorem that relates the pionic exchange current to the pion exchange potential.  To the best of our knowledge, there is no such theorem for the axial channel.}. We take this case as a direct confirmation of the IDD in the bsHLS Lagrangian.

It has been shown that one can also reproduce the long half-life by including chiral three-body forces that are subdominant to two-body forces in the chiral counting without invoking IDDs~\cite{holt-threebody,nocore}. One might argue that many-body forces in standard chiral perturbation theory are an alternative mechanism. But it is not quite right.  A better and more astute interpretation is that the two are essentially equivalent in the framework of effective field theory. That there is an overlap between the two interpretations, as far as the given process is concerned, was suggested in Ref.~\citen{holt-connection}.

\begin{figure}[ht]
\vskip 0.2cm
\begin{center}
\includegraphics[width=9.cm]{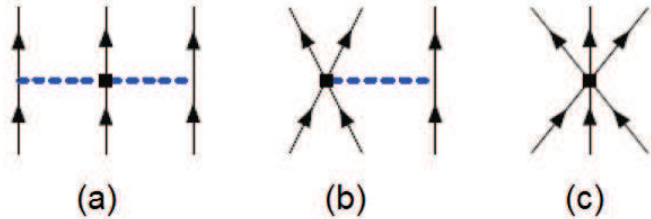}
\caption{The 3-body forces that figure in chiral perturbation calculations. The dotted line corresponds to $\pi$ exchange. The same three-body graphs with the dotted line representing $\pi$ and $\rho$ would figure in bsHLS.}
\label{3-body}
\end{center}
\end{figure}

 A simple way of understanding this connection between the two is as follows. Among the three three-body forces given in chiral perturbation theory,  Figs.~\ref{3-body} (a) and (b) of pionic range and  Fig.~\ref{3-body}(c) of zero-range, the latter can be thought, in the bsHLS approach, as pair-wise exchanges of $\omega$ mesons with the $\omega$s integrated out. The mass scale involved in this force is of the same scale from which the first decimation is done. Therefore that effect can be incorporated in the IDDs in the resulting two-body interactions, which is essentially what was done in Ref.~\citen{holt-c14}. Therefore if one were to include in the $V_{lowk}$ calculation the three-body forces, which is of course a legitimate procedure in the scheme, then with an appropriate IDD, the contact term would have to be given a smaller strength than what's used in Refs.~\citen{holt-threebody,nocore}. In addition, the pion in the figures (a) and (b) should be replaced by the $\pi$ and $\rho$ with their cancellation in the tensor force effective at increasing density properly taken into account. This would make their contributions to the tensor force strongly suppressed.

 It is possible that the pionic-range three-body forces, not captured in IDDs, could figure importantly in certain processes. An example is the oxygen anomaly~\cite{oxygen-anomaly}.  They can consistently, and should for generalization,  be included together with the $\rho$ exchange in the $V_{lowk}$ scheme.

\subsubsection{Nuclear matter}

The most recent review on applications to finite nuclei -- also to dense matter -- along the line described here is found in Ref.~\citen{ringdiagram}. Applied to nuclear matter, the theory gives the saturation properties of symmetric nuclear matter in a fair agreement with Nature, e.g., the binding energy $E_0/A=-15.1$ MeV, saturation density $n_{sat}=0.16$fm$^{-3}$ and the compression modulus $K=183.2$ MeV. The last is somewhat lower than the ``standard" value $\sim 220$ MeV, which reflects a softer EoS up to $n_{1/2}$ predicted by the theory. These results will be included in the figures given below for the EoS for dense matter for which R-II needs to be treated properly.
\subsection{Phenomena in R-II}
What makes the structure of R-II basically different from that of R-I is the density scaling of the vector mesons. While the properties of the baryons and the dilaton are primarily controlled by scale symmetry, the vector mesons are dictated by hidden local symmetry, particularly the vector manifestation which is scale-invariant. As stated above, with the vanishing (bilinear) quark condensate, the nucleon and dilaton masses are hardly affected by density up to near $n_c$. On the contrary, the $\rho$ mass -- and the hidden gauge coupling $g_\rho$ -- must go to zero near the VM fixed point. As for $\omega$, the $U(2)$ symmetry is found to break down badly in R-II, so it cannot be put together with $\rho$ in the flow to the M fixed point. There is a strong indication that the $\omega$ behaves totally differently from both the nucleon (and the dilaton) and the $\rho$.
\subsubsection{Symmetric nuclear matter and neutron matter}
\begin{figure}[thb]
\vskip 0.2cm
\begin{center}
\includegraphics[width=13cm]{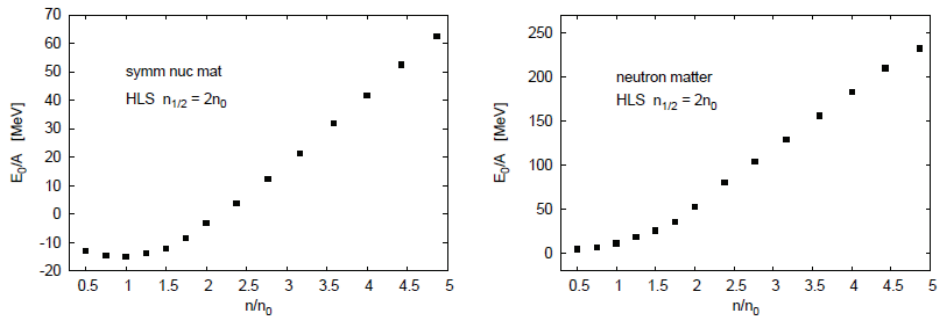}
\caption{Ground-state energy $E_0$ per nucleon
of symmetric nuclear matter (left panel) and neutron matter (right panel). The precise location of $n_{1/2}$ is not numerically significant. Here and in what follows it is taken to be $n_{1/2}=2n_0$.}
\label{E0}
\end{center}
\end{figure}

\begin{figure}[hb]
%\vskip 0.2cm
\begin{center}
\includegraphics[width=8.cm]{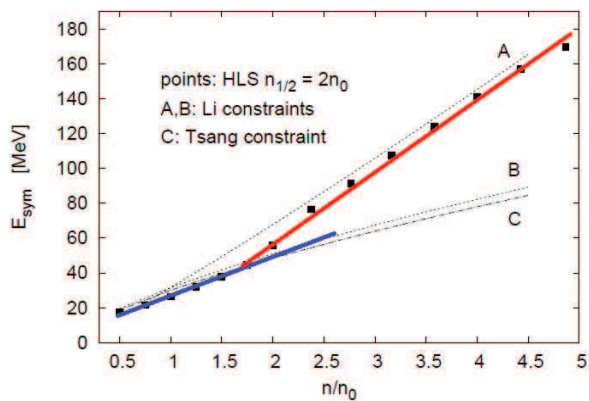}
\caption{The symmetry energy $E_{sym}$ predicted by the $V_{lowk}$ with bsHLS. The eye-ball slope change is indicated by the colored straight lines.}
\label{Esym}
\end{center}
\end{figure}

The energy per particle for symmetric nuclear matter and neutron matter is given in Fig.~\ref{E0}.  The notable feature of these results is that the symmetric matter is relatively soft, which is characterized by the somewhat low compression modulus $\sim 183$ MeV and the neutron matter is stiff to support $\sim 2$ solar-mass stars. This feature, consistent with heavy-ion data at low density and massive compact stars at high density,  can be seen in the symmetry energy Fig.~\ref{Esym}.

\subsubsection{Symmetry energy}
Let us return to the symmetry energy that illustrates the main thrust of our approach. We saw above that the collective-rotational correction to the energy of neutron-matter skyrmion crystal and the leading order term in $V_{lowk}$ in the closure approximation gave the same qualitative result. Going beyond the closure approximation, higher order correlations can be suitably taken into account in the ring-diagram approach to the $V_{lowk}$ calculation. The result taken from Ref.~\citen{PKLR} is given in Fig.~\ref{Esym}. The second decimation effectively ``smoothes" the cusp but exhibits the distinctly visible changeover from a soft EoS to a hard one in the region around $n_{1/2}$. While there is an apparent non-renormalizatpn of the tensor forces as mentioned above, the symmetry energy, dominated by the tensor force, receives non-negligible corrections from other components of the force and high-order correlations.  As mentioned, this is unlike the monopole matrix element involved in the shell evolution~\cite{otsuka} which zeros-in on what appears to be a fixed point-force.
\index{symmetry energy! skyrmion matter}
 \begin{figure}[ht]
%\vskip 0.2cm
\begin{center}
\includegraphics[width=8cm]{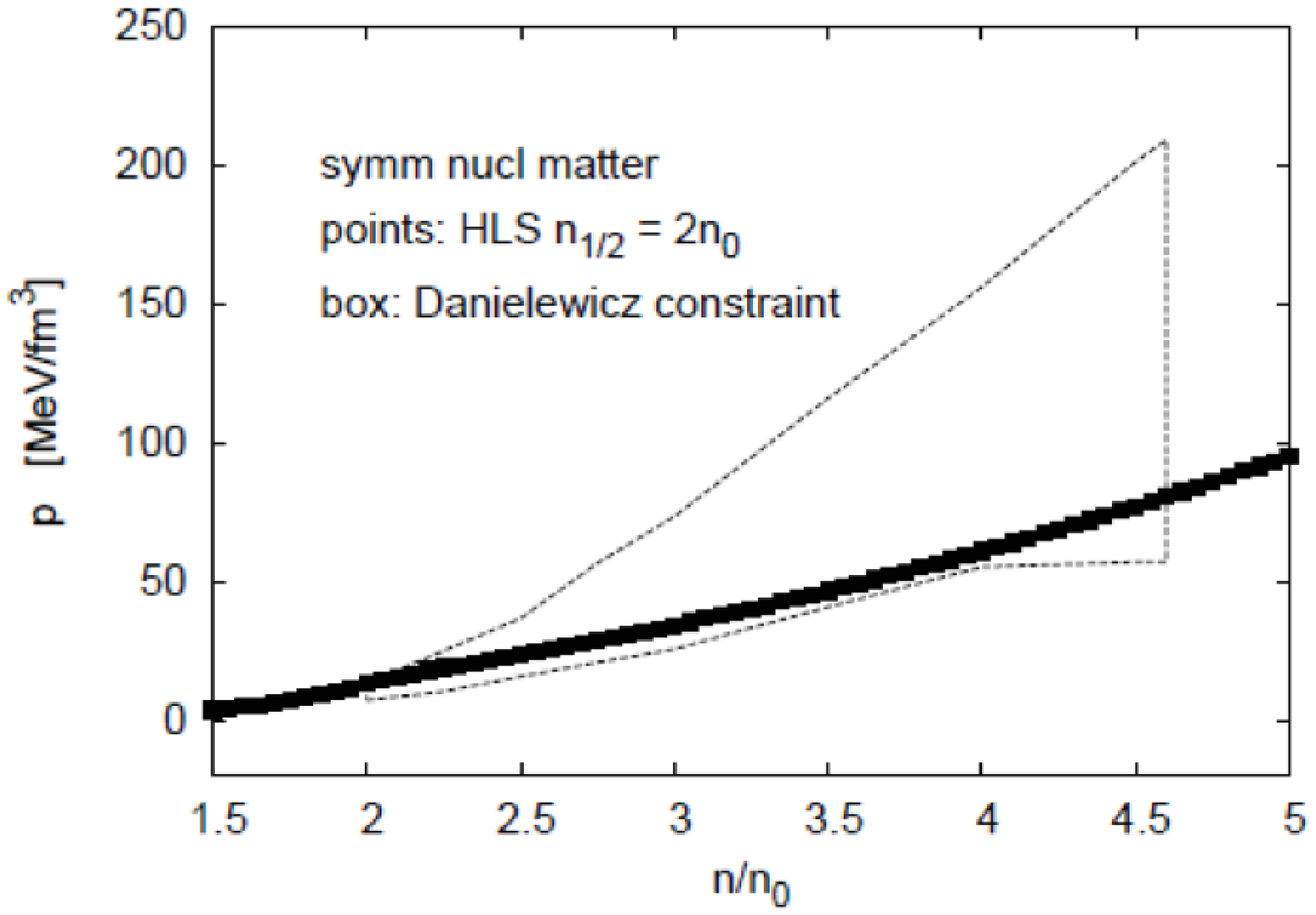}
\includegraphics[width=8cm]{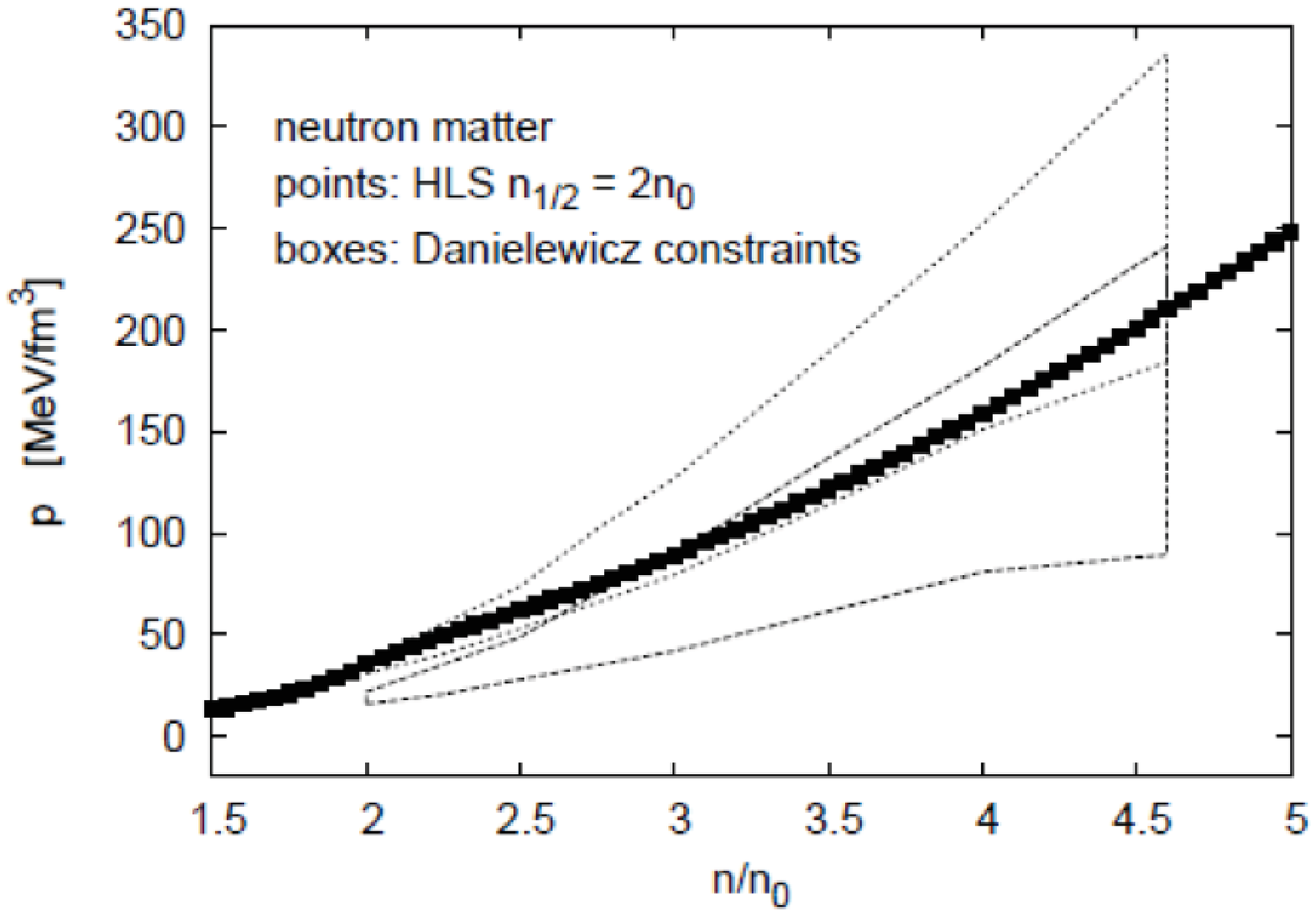}
\caption{Pressure vs. density for symmetric nuclear matter (upper panel) and for neutron matter (lower panel).}
\label{EoS}
\end{center}
\end{figure}

\subsubsection{Equation of State}

We have all the ingredients for calculating the EoS for compact stars. {The results obtained in Ref.~\citen{PKLR} are given in Fig.~\ref{EoS}} just to indicate that the results are compatible with the presently available experimental data.There is no doubt some room for refinement in the theory, so what's given here is not the final story. The EoS shows what was mentioned above, namely, that the EoS is relatively soft in nuclear matter and gets relatively harder in neutron matter, which implies that the same will be the case in compact-star matter in beta equilibrium.

 \begin{figure}[ht]
%\vskip 0.2cm
\begin{center}
\includegraphics[width=8cm]{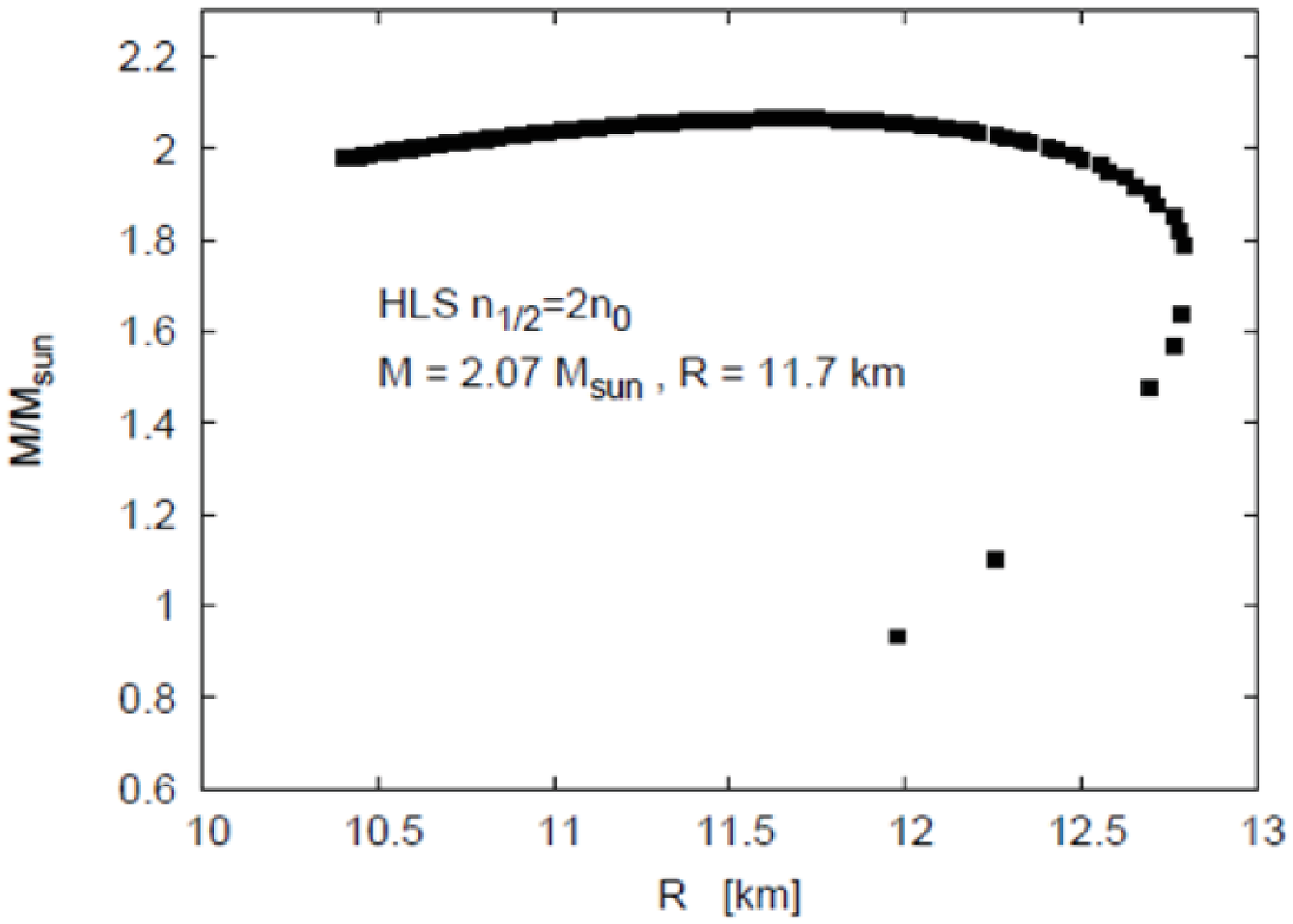}
\includegraphics[width=8cm]{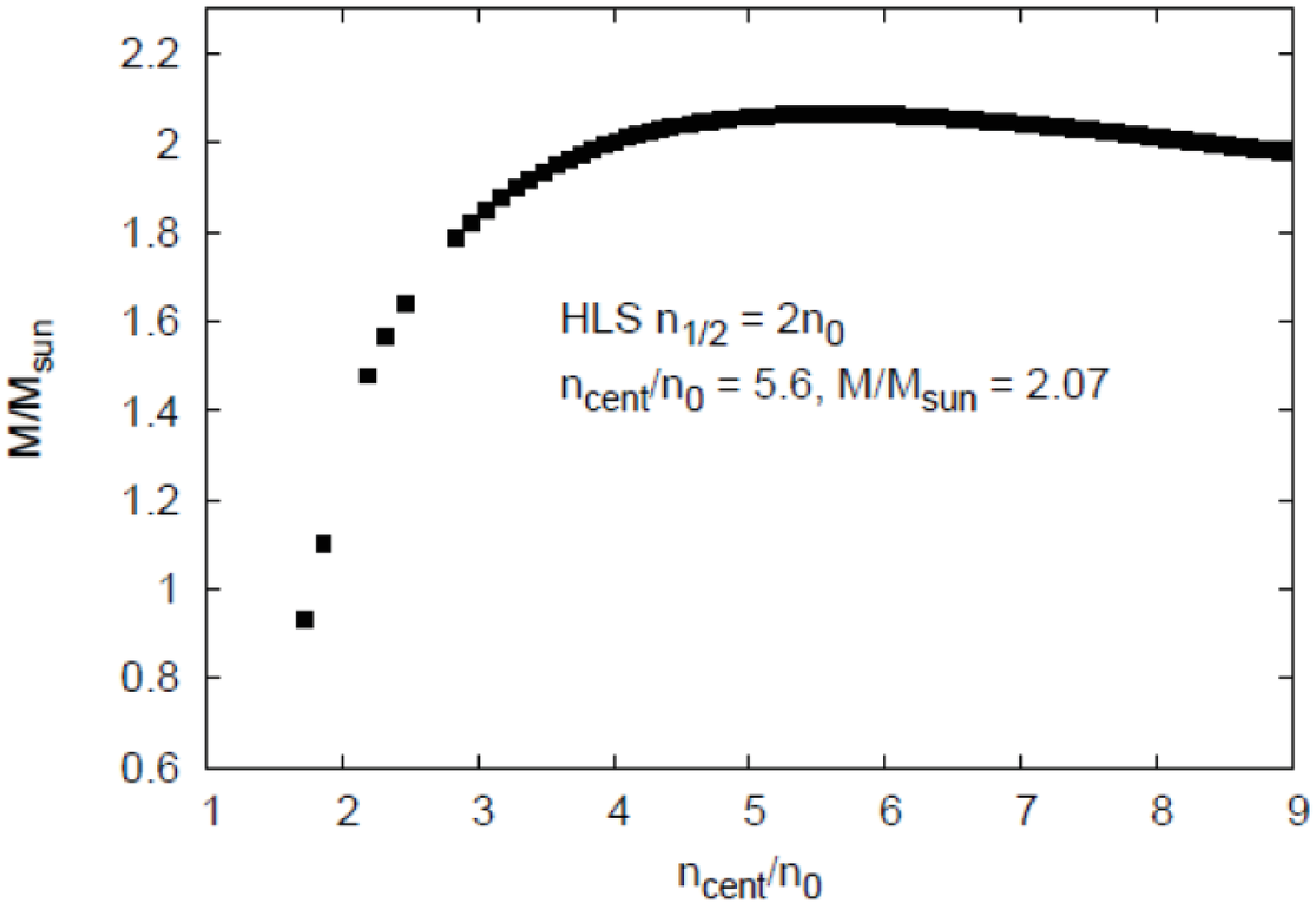}

\caption{Star mass vs.radius (upper panel) and star mass vs. central density (lower panel). }
\label{nstar}
\end{center}
\end{figure}

There have appeared several publications in the literature in which a smooth transition is effectuated from hadronic matter to quark matter at $\sim 2n_0$, with the consequent stiffening of the EoS at $n\gsim 2n_0$ as needed to support the observed $\sim 2$ solar mass stars.  Such a transition is made feasible by invoking strong vector interactions between quarks~\cite{hatsuda} or mediated by the quarkionic phase~\cite{quarkyonicstar}. While it has not been worked out, it is plausible and highly attractive that these mechanisms are dual (?) -- in the sense of Cheshire Cat -- to the half-skyrmion mechanism developed in this article. This needs to be worked out.

\subsection{Massive stars}

Finally the EoS so predicted is confronted with the structure of compact stars via the TOV equation. The results given  {in Ref.~\citen{PKLR}} are summarized in Fig.~\ref{nstar}.  Since the crust structure of the star is not taken into account,  $M$ vs. $R$  cannot be taken seriously for low-mass stars.  But the maximum-mass $2.07 M_\odot$ star with its corresponding radius $R=11.7$ km can be trusted. The central density of the maximum mass is found to be $5.6n_0$ and the sound velocity $v_s^2$ never exceeds $\sim $1/2 in units of $c$. It is consistent with the causality bound.

\section{Remarks}
We conclude by making a few comments on what we consider to be rather remarkable in the structure of dense baryonic matter. Combining the pseudo-NG dilaton to hidden gauge fields and exploiting what are thought to be robust topological properties of skyrmion crystal structure of dense matter, we arrive at rather striking and novel results up to date undiscovered. As skyrmions fractionize into half-skyrmions at $n\sim 2n_0$, there is an emergent parity doubling.  The quark condensate vanishes on average, but locally non-zero, supporting chiral density wave. There is pion with non-zero pion decay constant. There also emerges an equally important scale symmetry, hidden in the matter-free space, that gets intricately locked to chiral symmetry in dense matter giving rise to ``intrinisic density" dependence in the ``bare" parameters of the effective Lagrangian. The nucleon mass remains substantially ``non-melted," more or less density-independent within the range of density involved, with the bulk of the nucleon mass attributed to dilaton condensate and not to quark condensate. Both the $\rho$ meson and the $\omega$ meson, as hidden gauge fields, play crucial roles for all range of densities. At low densities, the tensor force, with an apparent scale invariance and an influence of the VM of the $\rho$ meson, impacts strongly on nuclear structure. At high densities, $n\gsim 2n_0$, the VM of the $\rho$ meson controls the symmetry energy, making it stiffer at higher density. There is an intricate interplay between the attraction due to  scalar meson (dilaton) exchange and the repulsion due to  $\omega$ meson exchange, leading to a mechanism that could, as suggested in Ref.~\citen{PKLR}, simply banish the strangeness degrees of freedom, i.e.,both hyperons and kaon condensations, beyond the maximum density relevant to compact stars, thereby resolving the ``strangeness problem" for the observed massive stars. In all ranges of density involved, there is no signal for explicit quark degrees of freedom, possibly reflecting the ``Cheshire Cat Principle" discussed in this volume by Nielsen and Zahed.

\subsection*{Acknowledgments}
We are grateful for discussions and collaborations with Tom Kuo, Won-Gi Paeng, Yongseok Oh and Byung-Yoon Park and for very helpful comments from Rod Crewther, Lewis Tunstall and Koichi Yamawaki on scale symmetry in both hadron and particle physics. This work was supported in part by the WCU project of Korean Ministry of Education, Science and Technology (R33-2008-000-10087-0). The work of MH was supported in part by the JSPS Grant-in-Aid for Scientific Research (C) N0.-24540266 and the work of YLM  by the National Science Foundation of China (NSFC) under Grant No.11475071 and the Seeds Funding of Jilin Univerity.

%\newpage

%\bibliographystyle{ws-rv-van}

\end{document}